\documentclass{aip-cp}

\usepackage[numbers,elide]{natbib}\usepackage{graphicx}
\usepackage{fancybox}
\usepackage{shortvrb}
\usepackage{braket}
\MakeShortVerb\|

\usepackage{listings}
\usepackage[usenames]{color}
\definecolor{codecolor}{gray}{.9}
\definecolor{rlcolor}{cmyk}{0,1,0,0}
\begin{document}

\title{Recent Theoretical Advances and Open Problems in Nuclear Cluster Physics}

\author[orsay,grenoble]{P. Schuck\corref{cor1}}
\affil[orsay]{Institut de Physique Nucl$\acute{e}$aire Univerisit$\acute{e}$ Paris-Sud,F-91406 Orsay Cedex, France}
\affil[grenoble]{LPMMC (UMR5493), Universit\'e Grenoble Alpes and CNRS, 25 rue des Martyrs, B.P. 166, 38042 Grenoble, France}
\corresp[cor1]{Corresponding author: schuck@ipno.in2p3.fr}

\maketitle

\begin{abstract}
This contribution gives a short review of recent theoretical advances in most topics of nuclear cluster physics concentrating, however, around $\alpha$ particle clustering. Along the route, the point of view will be critical mentioning not only progress but also failures and open problems.

\end{abstract}

\section{INTRODUCTION}
I was asked by the organisers to give an introductory talk on the
above given subject of ``Recent Theoretical Advances and Open Problems
in Nuclear Cluster Physics''. Nuclear cluster physics is an old
subject, almost as old as nuclear physics itself, see, e.g., reference
\cite{Teller} from 1936. Already at that time the $\alpha$ cluster was
the dominent cluster what is easy to understand, since it is the
smallest doubly magic nucleus and, thus, is almost as strongly bound
as the strongest bound nucleus which is $^{52}$Fe. Many relevant
experimental and theoretical works have appeared over the years. The
theoretical development of $\alpha$ cluster physics culminated in the
explanation of the by then still mysterious Hoyle state in $^{12}$C in
the early 70's in three independent works by H. Horiuchi, M. Kamimura
et al., and Uegaki et al. \cite{Horiuchi, Kamimura, Uegaki}. The Hoyle
state is the first excited $0^+$-state in the $^{12}$C nucleus, and in
those works it was described as a weakly bound $\alpha$ gas state
where the individual $\alpha$'s move mostly in relative
S-states. After these mile-stone works, it, curiously, became rather
silent in what concerns $\alpha$ clustering in nuclei. It was only in
2001 where a new idea concerning $\alpha$ gas states popped up and the
rate of publications on $\alpha$ clustering speeded up again quite
strongly. Namely, Tohsaki, Horiuchi, Schuck, and R\"opke (THSR)
proposed to interprete the Hoyle and also $\alpha$ gas-like states in
heavier nuclei as a condensate of $\alpha$ particles \cite{thsr}. This
idea was very successful and all properties of the well measured Hoyle
state could be reproduced without any adjustable parameters. Details
can be found, e.g., in recent review articles \cite{colloquium}.  This
idea of $\alpha$ condensation has now been strongly extended and also
other competing methods have been developed. For example, one can
describe now not only the Hoyle state but a whole family of Hoyle-like
states which can be considered as excited states of the Hoyle state
\cite{Funaki, Kimura}. But also ab initio methods like lattice quantum
Monte Carlo and shell model Monte Carlo approaches are being applied
to cluster states, as well as the symmetry adapted no core shell model
\cite {Meissner, Otsuka, Draayer}. I will report on these recent
developments in the main part of this contribution in section 2 and it
will actually take up a large part of this survey. In section 3, I
will highlight the situation in $^{16}$O. This will concern the 6-th
$0^+$ state and other cluster states in $^{16}$O. I will also report
on monopole transitions in connection with $\alpha$ clustering in
$^{16}$O. In section 4, I will outline recent efforts to describe
$\alpha$ clustering in nuclei with valence nucleons. Some part will be
devoted in section 5 to $\alpha$ condensation in infinite matter,
where, together with mean field studies in $^{16}$O, it will be shown
that $\alpha$ condensation is actually a Quantum Phase Transition
(QPT) where the density plays the role of the control
parameter. Finally in section 6 a new approach and insight into
spontaneous $\alpha$ decay of heavy nuclei will be given. In section
7, I will present the final conclusions and an outlook.\\
Of course my presentation will be biased towards my (our) own work but
I will try to include as much as possible also the work of other
colleagues.

\section{Various Aspects of the Hoyle state and its family}
The $\alpha$ condensate wave function is now known under the name of
THSR wave function. It is simply written out, e.g., for the three
$\alpha$ case, as

\begin{equation}
\Psi_{\mbox{THSR}} \propto {\mathcal A} \psi_1\psi_2\psi_3 \equiv {\mathcal A} \ket {B}
%{\mathcal A}e^{-(({\bf R}_1-{\bf X}_G)^2 + ({\bf R}_2-{\bf X}_G)^2 + ({\bf R}_3-{\bf X}_G)^2)/B^2}\phi_{\alpha_1}\phi_{\alpha_2}\phi_{\alpha_3}
\label{THSRwf}
\end{equation}

\noindent
with

\begin{equation}
\psi_i = e^{-(({\bf R}_i-{\bf X}_G)^2)/B^2}\phi_{\alpha_i}
\label{a-wf}
\end{equation}

\noindent
and

\begin{equation}
\phi_{\alpha_i} = e^{-\sum_{k<l}({\bf r}_{i,k}-{\bf r}_{i,l})^2/(8b^2)}
\label{int-a-wf}
\end{equation}

\begin{figure}
\parbox{5.5cm}{\vspace{0cm}
    \includegraphics[height=4.5cm]{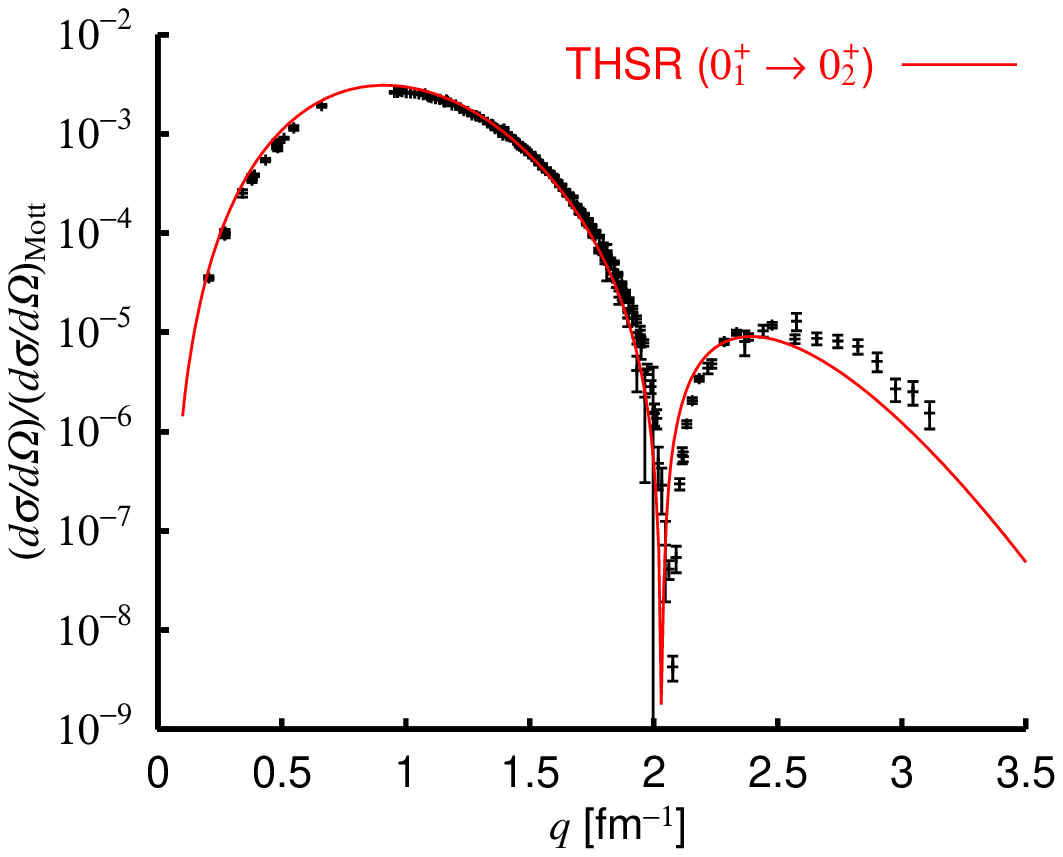}\hspace{0.5cm}}
   \parbox{5.5cm}{\vspace{0cm}
  \includegraphics[height=5cm,angle=-90]{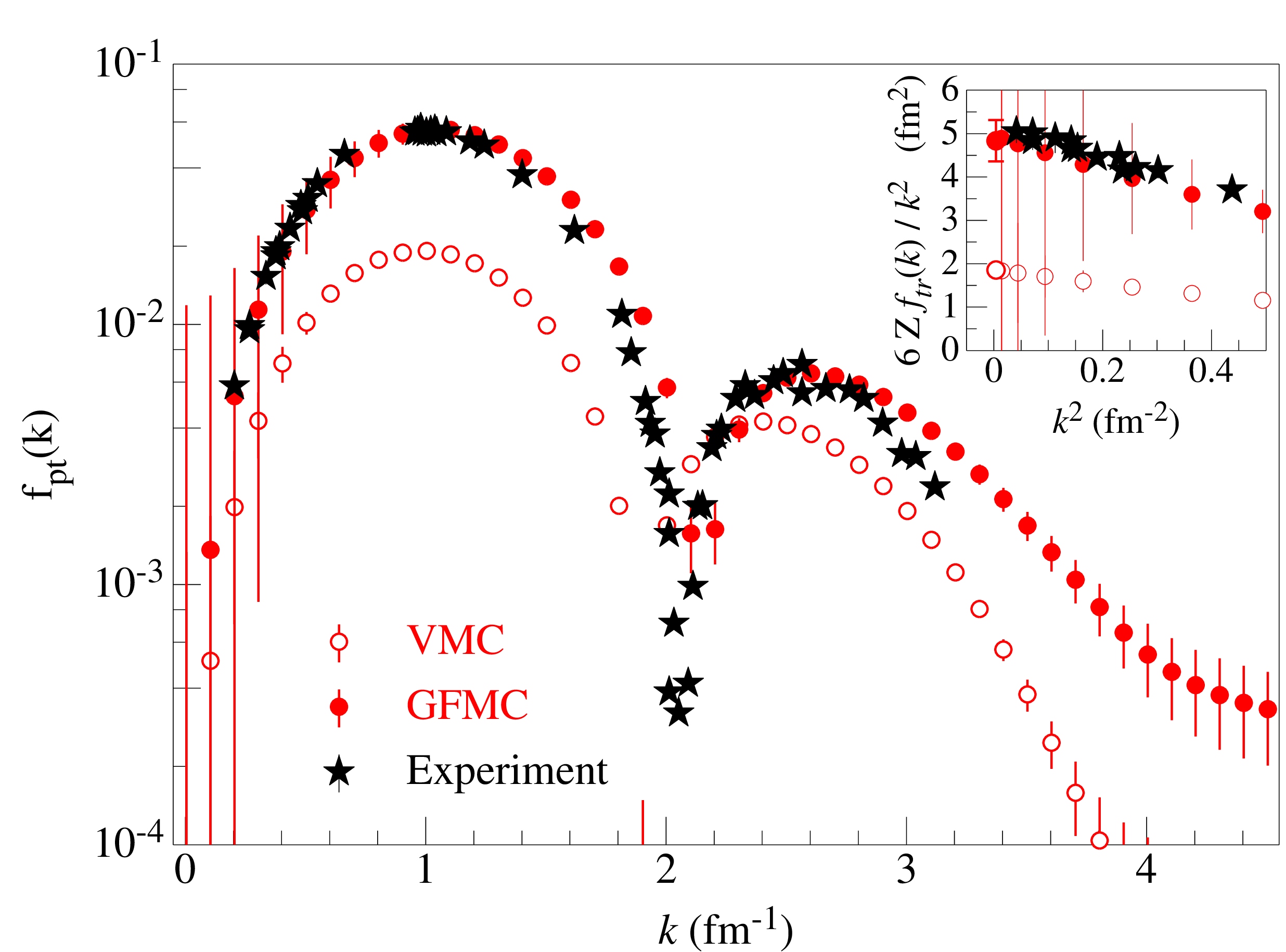}\hspace{0.5cm}}
\caption{\label{QMC} Inelastic form factors from THSR
  ~\cite{inelastic}, left panel, and from GFMC \cite{GFMC}, right
  panel. The THSR result cannot be distinguished from the one of
  \cite{Kamimura} on the scale of the figure. The open circles in the
  right panel correspond to the less accurate results from variational
  Monte Carlo (VMC).}
\end{figure}

\noindent
In Eq. (\ref{THSRwf}) the ${\bf R}_i$ are the c.o.m. coordinates of
$\alpha$ particle '$i$' and ${\bf X}_G$ is the total c.o.m. coordinate
of $^{12}$C. ${\mathcal A}$ is the antisymmetrizer of the twelve
nucleon wave function with $\phi_{\alpha_i}$ the intrinsic
translational invariant wave function of the $\alpha$-particle
'$i$'. The whole 12 nucleon wave function in Eq. (\ref{THSRwf}) is,
therefore, translationally invariant. Please note that we suppressed
the scalar spin-isospin part of the wave function. The special
Gaussian form given in Eqs.  (\ref{a-wf}), (\ref{int-a-wf}) was chosen
in \cite{thsr} to ease the variational calculation. The condensate
aspect lies in the fact that Eq. (\ref{THSRwf}) is a (antisymmetrized)
product of three times the same $\alpha$-particle wave function and
is, thus, analogous to a number projected BCS wave function in the
case of pairing.  This twelve nucleon wave function has two
variational parameters, $b$ and $B$. It possesses the remarkable
property that for $B=b$ it is a pure harmonic oscillator Slater
determinant (this aspect of Eq. (\ref{THSRwf}) is explained in
\cite{B-Bohr, Yam08}) whereas for $B \gg b$ the $\alpha$'s are at low
density so far apart from one another that the antisymmetrizer can be
dropped and, thus, Eq. (\ref{THSRwf}) becomes a simple product of
three $\alpha$ particles, all in identical 0S states, that is, a pure
condensate state. The minimization of the energy with a Hamiltonian
containing a nucleon-nucleon force determined earlier independently
\cite{Toh-force} allows to obtain a reasonable value for the ground
state energy of $ ^{12}$C. Variation of energy under the condition
that Eq. (\ref{THSRwf}) is orthogonal to the previously determined
ground state allows to calculate the first excited $0^+$ state, i.e.,
the Hoyle state. While the size of the individual $\alpha$ particles
remains very close to their free space value ($b \simeq$ 1.37 fm), the
variationally determined $B$ parameter takes on about three times this
value. It is important to mention right away that this so determined
THSR wave function has about 98 percent squared overlap with the one
of Kamimura {\it et al.} \cite{Kamimura} (and practically 100 percent
squared overlap with a slightly more general THSR wave function
superposing several $B$ values), see \cite{Yasuro1}.  Kamimura's et
al. wave function can, even to-day, after 40 years, be considered as
one of the most accurate approaches for the Hoyle state. In any case,
as in the work of Kamimura {\it et al.} \cite{Kamimura}, so does the
THSR approach reproduce very well all known experimental data about
the Hoyle state. This concerns for instance the inelastic form factor,
electromagnetic transition probability, and position of energy, see
for more details \cite{colloquium} and \cite{inelastic}. The inelastic
form factor is shown in Fig.\ref{QMC}. We see indeed very good
agreement with the experimental data. In the same figure we also show
recent results from Green's Function Monte Carlo (GFMC) calculations
\cite{GFMC}. They also reproduce the data very well. For the monopole
transition probability between Hoyle and ground state, even a 10 \%
better value is obtained. On the other hand the position of the Hoyle
state is not as accurate as the result with THSR. Let us mention that
also Chernykh {\it et al.} achieved to obtain with their Fermionic
Molecular Dynamics (FMD) approach a very reasonable inelastic form
factor \cite{Neff}. The related Antisymmetrised Molecular Dynamics
(AMD) method seems to perform slightly less well \cite{Enyo}.\\
Let us mention that at practically the same time as Kamimura et al.,
Uegaki {\it et al.} published a very similar paper leading to almost
identical results \cite{Uegaki}. In the following, we often will only
refer to Kamimura's work, since we were able to compare THSR and
Kamimura's wave functions numerically. However, all what we say below
about Kamimura's work should equally apply to the one of Uegaki.

%%%%%%%%%%%%%%%%%%%%%%%%%%%%%%%%%%%%%%%%%%%%%%%%%%%%%%%%%%%%%%%%%%%%%%%%%%
%%%%%%%%%%%%%%%%%%%%%%%%%%%%%%%%%%%%%%%%%%%%%%%%%%%%%%%%%%%%%%%%%%%%%%%%%%
Let us make at this point the following {\it very important
  observation}. Kamimura's wave function is, in principle much more
general than the THSR one. For instance Kamimura's wave function
contains for the Hoyle state also two $\alpha$, i.e., $^8$Be-like,
correlations. Nontheless, as we just mentioned, the fact that the
squared overlap between THSR and Kamimura wave functions is almost 100
\%, tells us that these extra $\alpha-\alpha$ correlations only play a
very minor role with respect to the conclusions obtained from the THSR
wave function where, besides correlations from antisymmetrisation,
such extra $\alpha-\alpha$ correlations are absent. In all what
follows, we should keep this fact in mind.

\subsection{The Condensation Aspect}

\begin{figure}
\includegraphics[height=6cm]{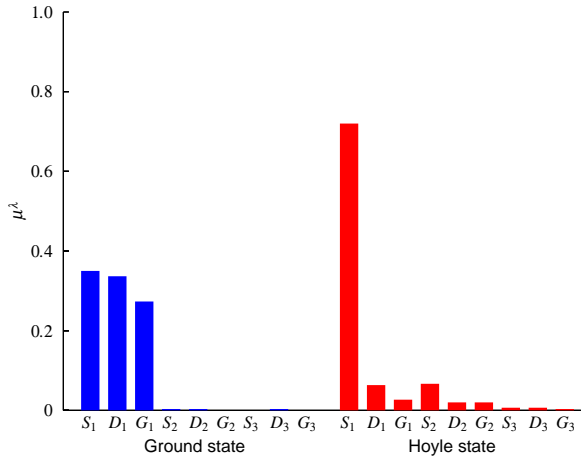}
  \caption{\label{occs-12C}
    $\alpha$ particle occupation numbers in the ground state
  (left) and in the Hoyle state (right) \cite{occ's}. The reader
  should make the following correspondencies for the spin-parity
  assignments: S$_1$ $\rightarrow$ 0S; D$_1$ $\rightarrow$ 0D; etc.}
\end{figure}

It is, of course very important to clear up the question whether the
Hoyle state is akin to a three $\alpha$ Bose-like condensate.  There
is a quantity which tells us directly whether the Hoyle state is close
to a three $\alpha$ condensate or not.  In \cite{Suzuki} Suzuki {\it
  et al.} evaluated the bosonic occupation numbers using a Gaussian
representation of the c.o.m. part $\chi$ of the RGM (Resonating Group
Method) wave function to calculate the single $\alpha$ particle
density matrix $\rho_{\alpha}({\bf R},{\bf R}')$ and diagonalized
it. The bosonic occupation numbers were also calculated by Yamada et
al. in \cite{occ's} using, however, the OCM approach \cite{ocm}. Both
calculations concluded that the three $\alpha$'s in the Hoyle state
occupy to about 70 percent the same 0S orbit whereas all other
occupations are down by at least a factor of ten, see
Fig.~\ref{occs-12C}. The $\alpha$ occupation numbers are a very
important quantity to decide whether we have an $\alpha$ condensate or
not. Unfortunately, very few calculations of this quantity
exist. Besides the two aforementioned ones, luckily there exists a
third one from where we can draw some general conclusions. Ishikawa
made for the Hoyle state a purely three boson model
\cite{Ishikawa}. Adjusting the force-parameters, he was able to
reproduce main data of the Hoyle state. For instance he also
calculated the bosonic occupation numbers. For the Hoyle state he
found $\sim$ 80 percent of condensation of the three bosons into the
0S state \cite{Ishi-private}. On the other hand he also found from his
full three body solution \cite{Ishikawa} that the $\alpha$'s are
moving with $ \sim$ 80 percent in relative S states. So there is a
direct relation between relative S-wave motion and S-wave
condensation. This correspondence is, of course, not extremely
surprising, rather confirms expectation. The point now is that there
exist about half a dozen Hoyle state calculations, beginning with the
one of Horiuchi \cite{Horiuchi} in 1974 where it was found that the
three $\alpha$'s move essentially in relative S-waves. So there is
little doubt that the Hoyle state can be interpreted with good
accuracy as a condensate of three $\alpha$ particles. A condensate is
an intrinsic state and, since the $\alpha$'s move in 0S states, the
intrinsic state is spherical.\\ Recently there have been several
experimental investigations which searched for a proof of the
existence of such a condensate in measuring the direct three $\alpha$
decay out of the Hoyle state \cite{3a-decay} with ever increasing
accuracy. For the moment the branching ratio with the two body decay,
$^8$Be + $\alpha$, is down to $10^{-4}$ with still no success. In a
remarkable work this number was confirmed theoretically by Ishikawa
\cite{Ishikawa}, see also \cite{Bonasera}. However, absence of
measurement of direct three $\alpha$ decay does not mean absence of
$\alpha$ condensate. On their decay, the three $\alpha$'s have to
tunnel through the Coulomb barrier and are distorted by the final
state Coulomb interaction so that it may be extremely difficult to get
a clear signal from the internal $\alpha$ structure of the Hoyle state
from the direct 3-$\alpha$ decay. However, this does not mean that the
experimental effort should be stopped! It would, indeed, be very nice
if a direct three $\alpha$ decay could be found.  In this respect it
is interesting to note that in \cite{Bonasera} it is found that the
branching ratio for direct three $\alpha$ decay increases very much
(up to 10\%) if a condensate state appears at around 10 MeV. The 3rd
$0^+$ could be a candidate, see next section.

%%%%%%%%%%%%%%%%%%%%%%%%%%%%%%%%%%%%%%%%%%%%%%%%%%%%%%%%%%%%%5
%%%%%%%%%%%%%%%%%%%%%%%%%%%%%%%%%%%%%%%%%%%%%%%%%%%%%%%%%%%%%%%%5

\subsection{The Hoyle Family of States}

The Hoyle state seen as a Bose condensate can be interpreted as the
ground state of a new tower of excited states. And, indeed, there
are. The $0^+_3$ and $0^+_4$ states at 10.08 MeV and 11.87 MeV have
been identified as a breathing mode and a bent arm excitation of the
Hoyle state, respectively, see the early work of Kato {\it et al.}
employing the OCM method \cite{Kato}, and also the more recent work by
Bo Zhou {\it et al.} \cite{Bo1}. One may speculate that the $0^+_3$
state is a state where the three $\alpha$'s move in relative 1S orbits
and, thus, can be qualified to be again a Bose condensate, however,
one with a higher nodal structure. More work has to be investigated on
this exciting aspect before firm conclusions can be reached (I am very
greatful to Bo Zhou for a discussion on this point).  The direct three
$\alpha$ decay out of breathing Hoyle state may be particularly
interesting. What happens to an excited Bose condensate when it decays
democratically can be seen in \cite{explosion} (I am greatful to
M. Freer for having informed me about this reference).\\
In the theoretical description of the excited Hoyle states, there have
recently been two advances. First Y. Funaki extended the THSR wave
function including one more width parameter \cite{Funaki}, see also
\cite{Bo2}. This allows to describe correlations between two
$\alpha$'s. For excited Hoyle states, this is important. Second, very
recently, M. Kimura et al. \cite{Kimura} proposed a novel method
called 'Real-Time Evolution Method' (REM).  The method utilizes the
equation-of-motion of the Gaussian wave packets to generate the basis
wave functions having various three $\alpha$ cluster configurations.
The generated basis wave functions are superposed to diagonalize the
Hamiltonian. In other words, this method uses the real time as the
generator coordinate. The application to the 3 $\alpha$ system as a
benchmark shows that the new method works efficiently in selecting
relevant $\alpha$ configurations yielding results consistent with or
better than the other cluster models. Also the inelastic form factor
to the Hoyle state comes out quite reasonable \cite{Kimura-p}. The
structure of the excited $0^+$ and $1^-$ states in $^{12}$C is also
elucidated. For instance it was found that the transition probability
between the Hoyle state and the $1^-$ state has the huge value of 8.8
Weisskopf units, see Fig.\ref{Kimura2}. It, therefore, can be
speculated that the $1^-$ state corresponds to lifting an $\alpha$
particle from the Hoyle state into the P-shell. Also the radius of
this $1^-$ state is found to be enormous. It would be very important
that this new finding of this $1_1^-$ state being an $\alpha$ gas
state be confirmed. It also may then need a re-interpretation of the
5th $0^+$ state in $^{16}$O which is believed to consist of an
$\alpha$ orbiting around the $1^-$ state in $^{12}$C, see also section
3.

%%%%%%%%%%%%%%%%%%%%%%%%%%%%%%%%%%%%%%%%%%%%%%%%%%%%%%%%%%%%%%%%%%%%%%%%%%
\begin{figure}
  \includegraphics[width=15cm]{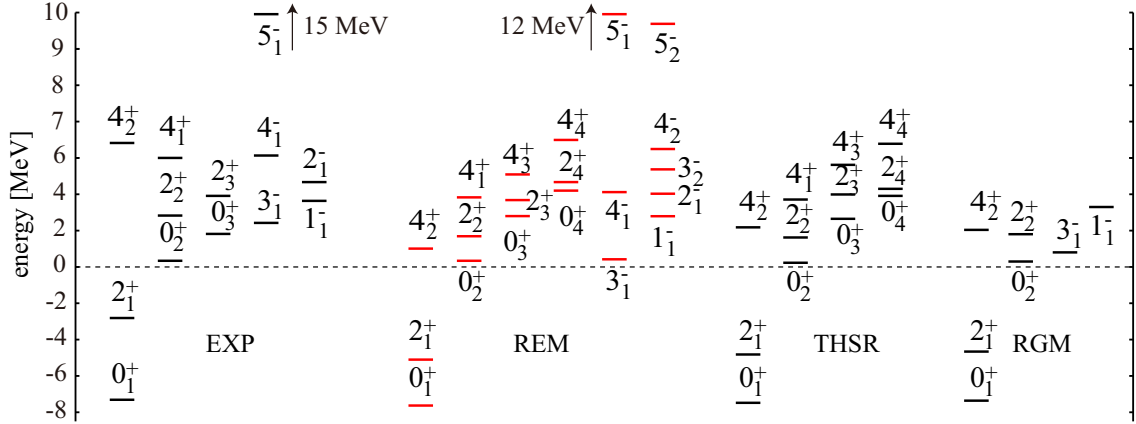}
  \caption{\label{Kimura1} The $\alpha$ gas spectrum of 12C from three
    different methods: second group of states, REM as described in the
    text; third group, extended THSR by Funaki {\it et al.}
    \cite{Funaki}; the early but pioneering results by Kamimura et
         {\it al.} with the Resonating Group method
         \cite{Kamimura}. The first group shows the experimental
         spectrum. From \cite{Kimura}.}
\end{figure}
%%%%%%%%%%%%%%%%%%%%%%%%%%%%%%%%%%%%%%%%%%%%%%%%%%%%%%%%%%%%%%%%%%%%%%

%%%%%%%%%%%%%%%%%%%%%%%%%%%%%%%%%%%%%%%%%%%%%%%%%%%%%%%%%%%%%%%%%%%%%%
\begin{figure}
\includegraphics[width=10cm]{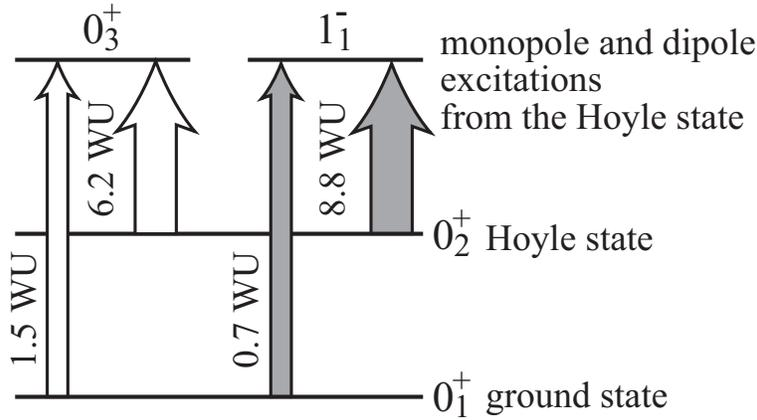}
  \caption{\label{Kimura2} Transition probability between Hoyle state
    and the first $1^-$ state in $^{12}$C. From \cite{Kimura}}
\end{figure}
%%%%%%%%%%%%%%%%%%%%%%%%%%%%%%%%%%%%%%%%%%%%%%%%%%%%%%%%%%%%%%%%%%%%%
%%%%%%%%%%%%%%%%%%%%%%%%%%%%%%%%%%%%%%%%%%%%%%%%%%%%%%%%%%%%%%%%%%%%%%
\begin{figure}
  \includegraphics[width=10cm]{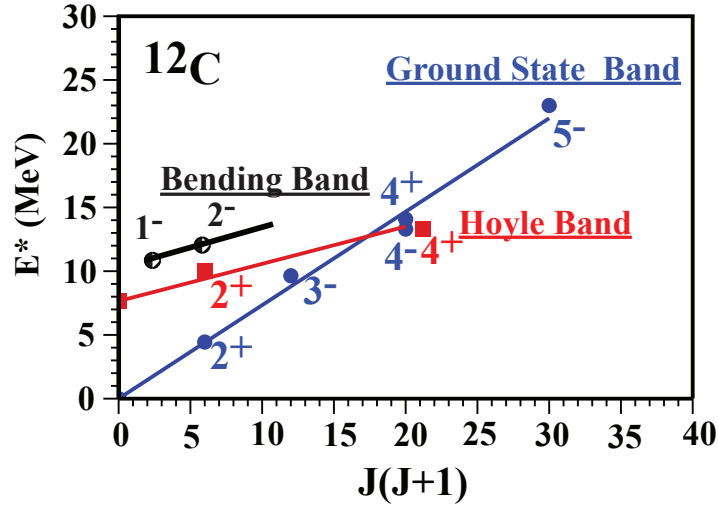}
  \caption{\label{Freer1} The rotational band structure obtained from
    the spinning triangle model in comparison with experiment. From
    \cite{Gai}. Figure greatfully supplied by M. Freer.}
  \end{figure}
%%%%%%%%%%%%%%%%%%%%%%%%%%%%%%%%%%%%%%%%%%%%%%%%%%%%%%%%%%%%%%%%%%%%%%

%%%%%%%%%%%%%%%%%%%%%%%%%%%%%%%%%%%%%%%%%%%%%%%%%%%%%%%%%%%%%%%%%%%%%
\begin{figure}
  \includegraphics[width=6cm]{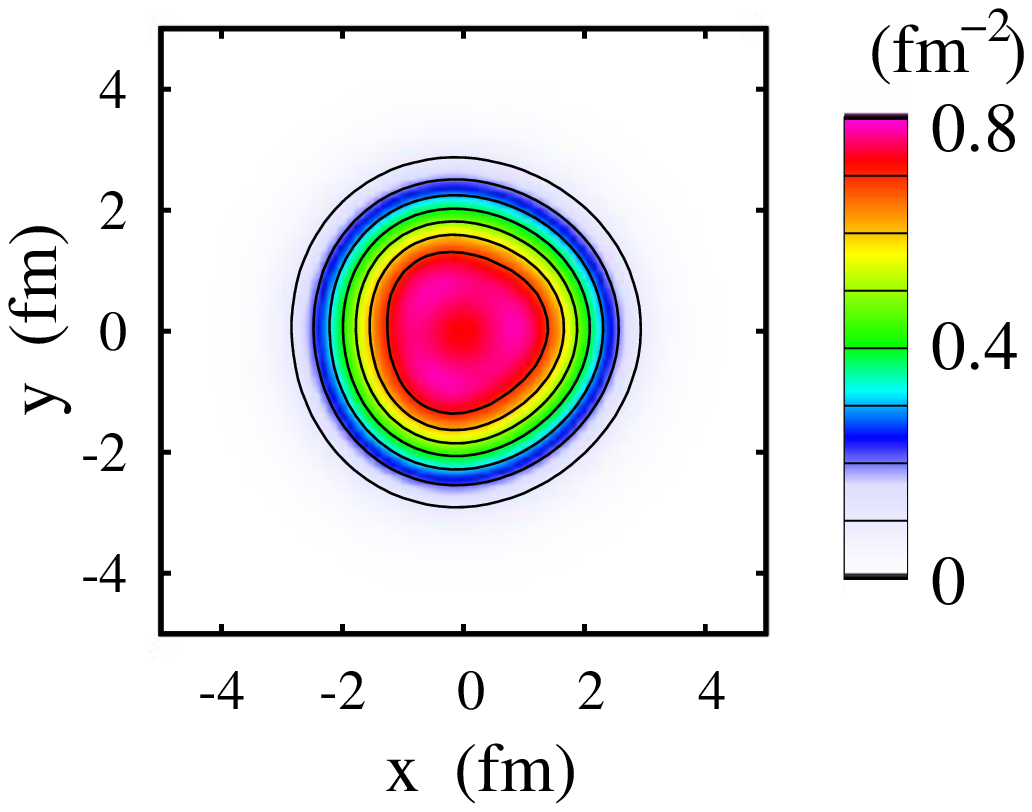}
  \caption{\label{Yoshiko}The triangular shape obtained from an AMD
    calculation (courtesey Y. Kanada-En'yo).}
\end{figure}
%%%%%%%%%%%%%%%%%%%%%%%%%%%%%%%%%%%%%%%%%%%%%%%%%%%%%%%%%%%%%%%%%%%%%

\subsection{Rotational bands in $^{12}$C}

Recently some algebraic models were published which seem to explain
very nicely the ground state rotational band and, to some extent, also
a rotational band with the Hoyle state as band head \cite{Gai}. This
model supposes that the ground state of $^{12}$C has triangular shape
and it, therefore, describes the spectrum of a spinning triangle. The
result is shown in Fig.\ref{Freer1}. At first, the reproduction of the
experimental ground state band seems very convincing. However, there
may be some caveats to this nice picture. First Peter Hess showed
\cite{Hess} that one cannot neglect the Pauli principle as done in
\cite{Gai}. Also Cseh \cite{Cseh} invented a successful model where he
groups the positive and negative parity states of the band in two
separate bands, respectively. Second, the indication of a triangularly
shaped ground state of $^{12}$C is rather weak. Indeed, a calculation
by Kanada-En'yo shows in Fig.\ref{Yoshiko} a slight triangular
shape. This figure is obtained from an AMD calculation and has been
sent to the author privately. In principle it is a pure mean field
calculation and, therefore, can be qualified as a (weak) spontaneous
symmetry breaking.

In \cite{Yoshiko} one sees similar figures which have been obtained
from an AMD calculation with angular momentum projection {\it before}
variation and, thus, the triangular shapes become more pronounced. I
have asked several of my colleagues to perform a pure state of the art
mean field (Slater determinant, relativistic and non-relativistic)
calculation without projection. Never anything else than a perfectly
round oblate shape was obtained. It makes a fundamental difference
whether a deformation is spontaneously obtained in mean field (as,
e.g., with Rare Earth nuclei), or whether a deformation is only
obtained with projection before variation. In the latter case the
systems only has shape fluctuations but no permanent deformation. So,
we see, the notion of a spinning triangle may be on not so safe
grounds. Concerning the so-called Hoyle band also shown in
Fig.\ref{Freer1} there exist even doubts that one can call this a band
at all. The trouble comes from the fact that it was shown that the
transition probability between the second $2^+$ and $0^+_2$ and
$0^+_3$ states are of similar magnitude \cite{Funaki}. So no well
defined band head exists. The $2^+_2$ state can be interpreted as a
quadrupole excitation of the Hoyle state \cite{Funaki}.

\subsection{Ab Initio Methods}

\begin{figure}[htbp]
    \includegraphics[width=10cm]{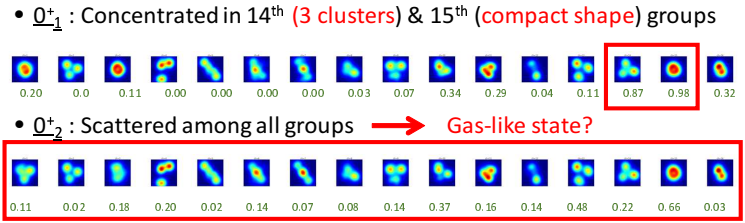}
\caption{\label{Otsuka1} The different configurations contributing to
  the Hoyle state obtained from the no-core shell model (MCSM)
  approach. Notice the numerous significant contributions to the Hoyle
  state whereas only two configurations contribute with large weights
  to the ground state. (Figure greatfully supplied by Takashi Abe).}
\end{figure}

Recently, there also has been great progress in describing $\alpha$
gas states with ab initio techniques. The most advanced is certaily
the GFMC method \cite{GFMC} with its very good reproduction of the
inealstic form factor to the Hoyle state which was already discussed
at the beginning of this section.\\
Among the shell model approaches the best achievement seems to come
from the no-core Monte Carlo shell model (MCSM) approach by Otsuka et
al. \cite{Otsuka}. In Fig.\ref{Otsuka1} we show the result of their
calculation for the Hoyle state. We see that the ground state is
concentrated in only a few configurations while the Hoyle state shows
many significant contributions. This is a clear sign that the system
wants to reproduce the relative S-wave motion of the $\alpha$'s. So
far no inelastic form factor has been calculated but the monopole
transition probability comes out quite reasonable. Also the position
is reasonably accurate. However, to be totally convinced of the
approach the inelastic form factor is needed.\\

%%%%%%%%%%%%%%%%%%%%%%%%%%%%%%%%%%%%%%%%%%%%%%%%%%%%%%%%%%%%%%%%%%%%%%
%%%%%%%%%%%%%%%%%%%%%%%%%%%%%%%%%%%%%%%%%%%%%%%%%%%%%%%%%%%%%%%%%%%%%%
There exists also the symmetry-adapted no-core shell model (NCSM) by
J. Draayer et al. \cite{Draayer}. Quite good progress has recently
been achieved for the position of the Hoyle state. However, no
transition probability or inelastic form factor have been
calculated. As said above, without those ingredients, it is difficult
to judge the value of the approach.\\
%%%%%%%%%%%%%%%%%%%%%%%%%%%%%%%%%%%%%%%%%%%%%%%%%%%%%%%%%%%%%%%%%%%%
\begin{figure}
  \includegraphics[width=6cm]{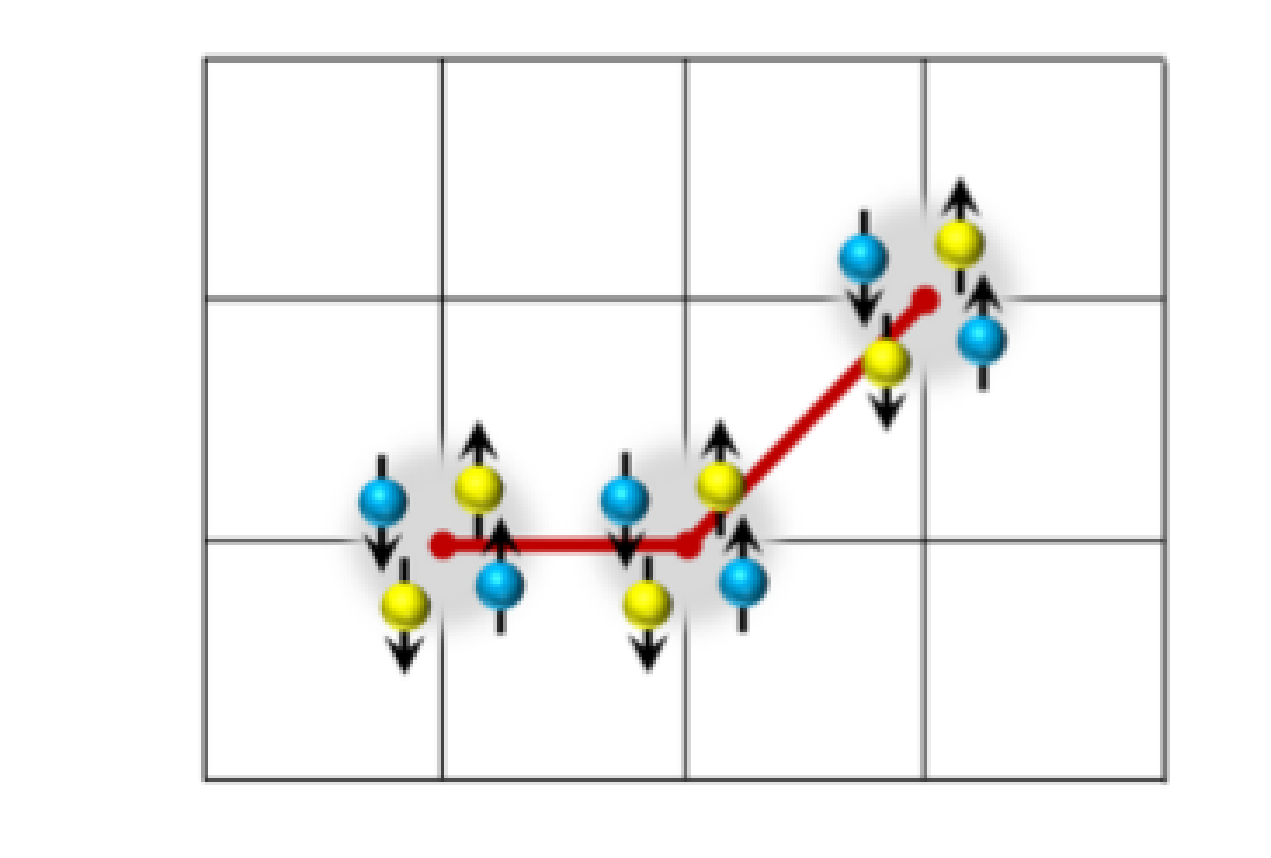}
  \caption{\label{Meissner1} Dominant Hoyle configuration obtained
    with the lattice QMC-EFT approach. From \cite{Meissner}. Figure
    greatfully provided by E. Epelbaum.}
  \end{figure}
%%%%%%%%%%%%%%%%%%%%%%%%%%%%%%%%%%%%%%%%%%%%%%%%%%%%%%%%%%%%%%%%%%%
Quite some publicity has been made recently around the ab initio
lattice quantum Monte Carlo (QMC) effective field theory (EFT)
approach by Epelbaum {\it et al.} \cite {Meissner}. In
Fig. \ref{Meissner1} we show the configuration of the Hoyle state
obtained by these authors.  As one sees the configuration of the Hoyle
state is quite opposite of what we have concluded above, namely that
the Hoyle state can be described to a good approximation as a state
where the center of mass motion of the three $\alpha$ particles is
condensed into the lowest S-wave. This bent arm configuration can at
best be attributed to the fourth $0^+$ state, see section II.B. It is
impossible that a configuration of relative S-waves acquires such a
shape. Nevertheless, the spectrum is surprisingly well reproduced from
this ab initio approach starting from the Chiral effective Lagrangian
which fits the NN scattering data. On the other hand neither inelastic
form factor nor the transition probability have been attempted to be
calculated. Even the radius, see below, completely fails since it is
barely larger than the one of the ground state. This may not be
astonishing, since the mesh size used is with 2 fm still much too
large. However, to the defense of the authors, one should mention that
the radius has only been calculated to LO whereas the energies are
evaluated to NNLO. It seems that Lattice QMC-EFT has still quite some
way to go before it can compete with the other methods.

\subsection{The radius of the Hoyle and Hoyle like states}

Since the first calculations of the Hoyle state, its radius has been
subject to discussion. Most theoretical calculations now agree that
its rms radius lies somewhere between 3.4 fm and 3.8 fm (ground state:
2.4 fm). With the THSR approach it was found that the magnitude of the
inelastic form factor is very sensitive to the extension of the Hoyle
state \cite{inelastic}. So the parameter free reproduction of the
inelastic form factor was taken as good sign for an enlarged volume of
the Hoyle state. If one believes into the condensation picture, this
also seems natural from the physical point of view, since only in an
enlarged volume the $\alpha$'s can be born out and acquire an almost
free gas motion. On the other hand experimentalists struggled to
confirm this hypothesis. For instance Ogloblin et al. \cite{Ogloblin}
found only very little radius enhancement of the Hoyle state. It is
only very recently that Makoto Ito \cite{Ito} via a very careful
coupled channel analysis could conclude from his measurement that the
second $2^+$ state in $^{12}$C has an enhanced radius of $\sim$ 3.4
fm, that is 1fm more than the ground state. The second $2^+$ state has
clearly been identified only since short to belong to the Hoyle family
\cite{Itoh, Gai2} and one can quite safely extrapolate from this study
that also the Hoyle state itself has a radius of similar extended
magnitude. Several authors found a strong transition probability
between the Hoyle state and the second $2^+$ state. So the issue of
$\alpha$ gas states having a volume 3-4 time larger than the ground
state seems finally to be settled.

\section{The Situation in $^{16}$O}

%%%%%%%%%%%%%%%%%%%%%%%%%%%%%%%%%%%%%%%%%%%%%%%%%%%%%%%%%%%%%%%%%%5
\begin{figure}
  \includegraphics[width=10cm]{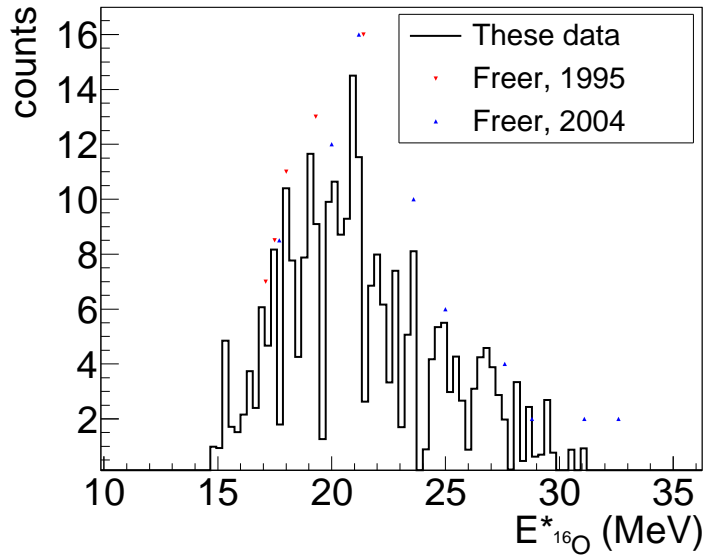}
  \caption{\label{Barbui} The excitation spectrum of $^{16}$O obtained
    from the $^{20}$Ne on $\alpha$ inverse kinematic reaction, see
    \cite{Natowitz1} . Clearly a state at $\sim$ 15.2 MeV obtained
    from a two $^8$Be coincidence measurement together with an
    $\alpha$ plus Hoyle state coincidence can be identified. The other
    resonances coincide with the ones measured earlier By M. Freer
    {\it et al.} \cite{Freer4}. Figure greatfully supplied by
    M. Barbui.}
\end{figure}
%%%%%%%%%%%%%%%%%%%%%%%%%%%%%%%%%%%%%%%%%%%%%%%%%%%%%%%%%%%%%%%%%%%

The situation in $^{16}$O with respect to $\alpha$ gas states has also
evolved quite a bit (see, e.g., for the theory side
\cite{Yasuro2}). But, unfortunately, for instance the experimental
situation is by far not so satisfying as in $^{12}$C. The 6-th $0^+$
state at 15.1 MeV which is 700 keV above the four $\alpha$ threshold
is suspected to be the analogue to the Hoyle state
\cite{Yasuro2}. However, unfortunately, the inelastic form factor
which is of so precious information for the Hoyle state has still not
been measured. Very recently, even more trouble came up concerning
this state. Careful measurements by Freer {\it et al.} \cite{Freer2}
have revealed that very close to this 6-th $0^+$ state is located
another state with a spin different from $0^+$ which disturbs the
analysis and interpretation of the $0_6^+$ state. On the other hand,
recent measurements by M. Barbui {\it et al.} \cite{Natowitz1} show in
the missing mass spectrum obtained from the $^{20}$Ne $+ \alpha$
inverse kinematics reaction a clear peak at 15.2 MeV which might be
the searched for four $\alpha$ gas state. This very promising result
(courtesey M. Barbui) is shown in Fig.\ref{Barbui}.  The peak at 15.2
MeV has been obtained from de-exciting $^{16}$O nucleus measuring in
coincidence two $^8$Be or one $\alpha$ plus a $^{12}$C in the Hoyle
state. It would be desirable to have the spin asignment of this
state. The other peaks are known resonances agreeing with earlier
measurements by other authors, see, e.g., \cite{Freer4}. The decay
into two $^8$Be is particularly interesting because $^8$Be is, of
course also a boson and this may indicate the condensation of two
$^8$Be's. Of course, the condensation of two $^8$Be's or of four
$\alpha$ can barely be distinguished as we have already discussed for
the case of the Hoyle state concerning the $\sim $ 100 $\%$ square
overlap of Kamimura and THSR wave functions. However, on the
theoretical level the question whether a Bose gas condenses as singles
or as molecules is a very interesting one \cite{Noz}.\\

%%%%%%%%%%%%%%%%%%%%%%%%%%%%%%%%%%%%%%%%%%%%%%%%%%%%%%%%%%%%%55
\begin{figure}
\includegraphics[width=6cm]{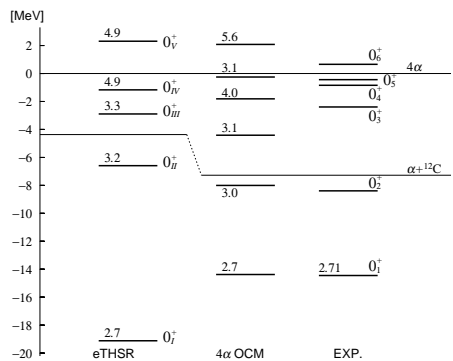}
\caption{\label{Funaki3} The $0^+$ spectrum of $^{16}$O calculated
  with an extended THSR wave function and with the OCM method. The
  right column shows the experimental spectrum. From \cite{Funaki3}}
\end{figure}
%%%%%%%%%%%%%%%%%%%%%%%%%%%%%%%%%%%%%%%%%%%%%%%%%%%%%%%%%%%%%%5

%%%%%%%%%%%%%%%%%%%%%%%%%%%%%%%%%%%%%%%%%%%%%%%%%%%%%%%%%%%%%%%%%%
\begin{figure}
\includegraphics[width=6cm]{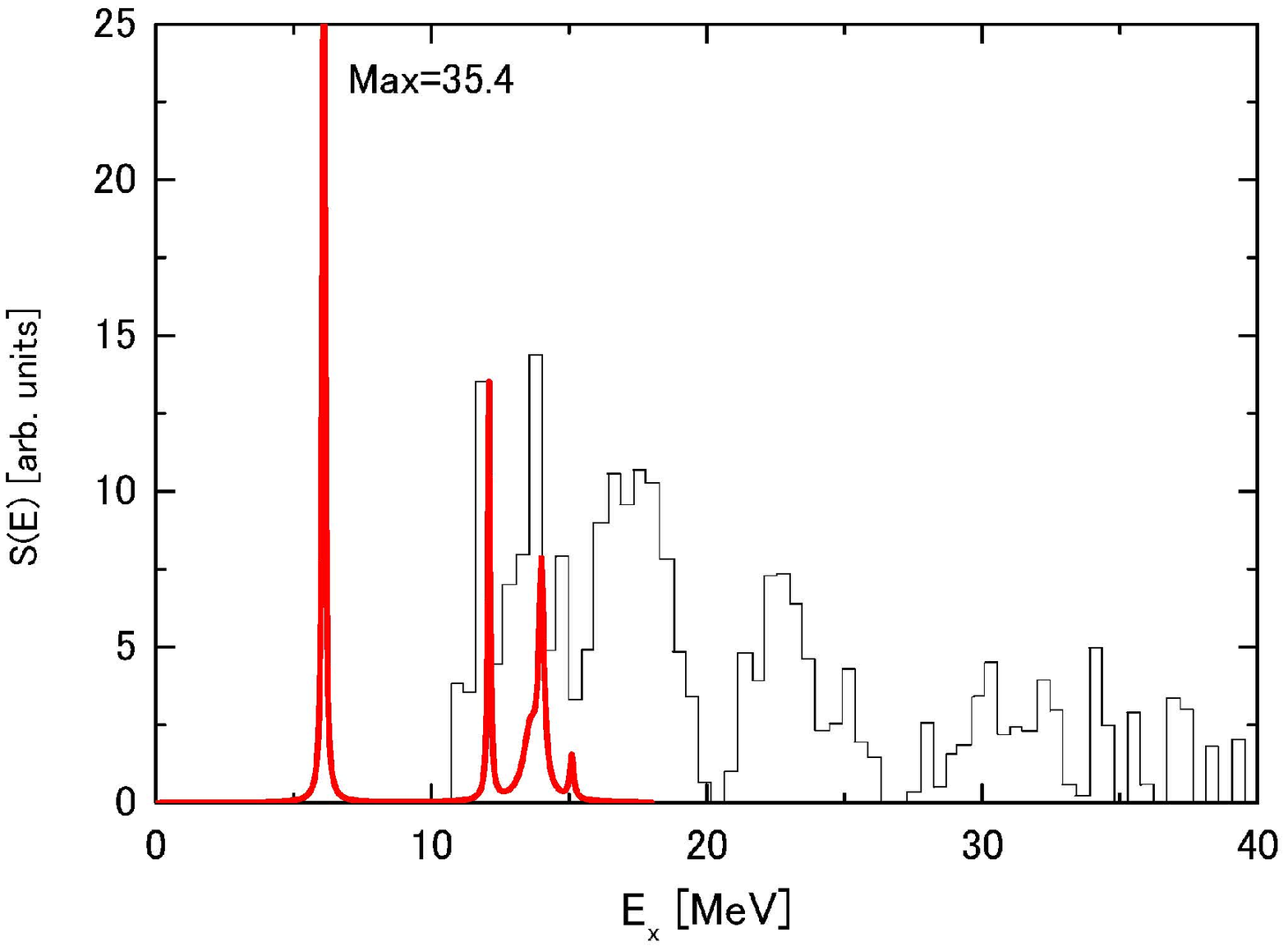}  
\includegraphics[width=6cm]{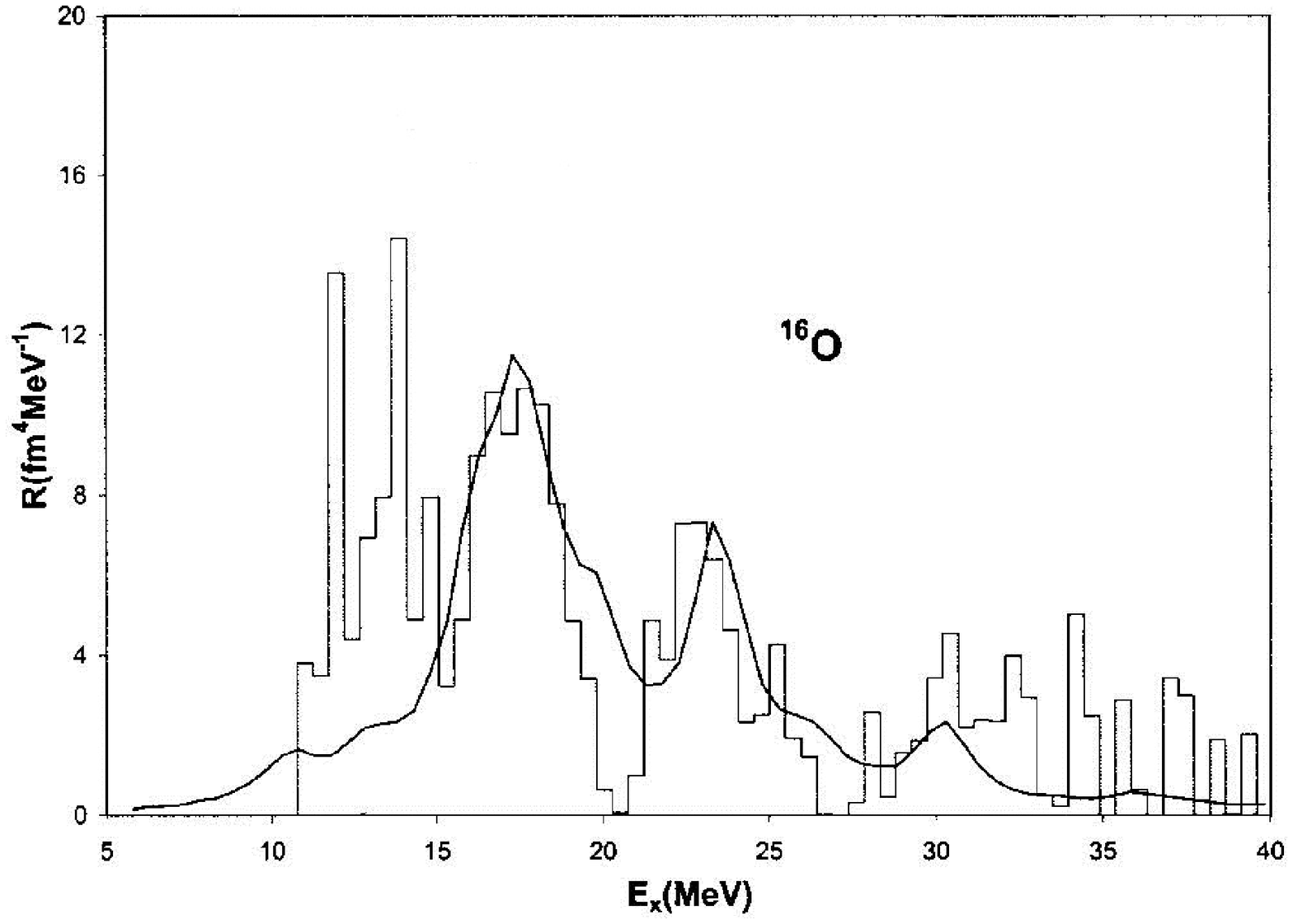}
\caption{\label{Yamada1,2} The monopole transition strength in
  $^{16}$O is shown as a function of excitation energy. We see in the
  left panel that the cluster states reproduce the low lying part of
  the spectrum whereas the high lying part can be approximated by the
  traditional RPA approach, right panel. From \cite{Yamada1}.}
\end{figure}
%%%%%%%%%%%%%%%%%%%%%%%%%%%%%%%%%%%%%%%%%%%%%%%%%%%%%%%%%%%%%%%

On the theory side things have progressed significantly. Funaki
\cite{Funaki3} has made a quite complete study of the $0^+$ spectrum
in $^{16}$O. It is shown in Fig.\ref{Funaki3}. He used an extended
THSR wave function where he considers one $\alpha$ on top of the Hoyle
state. Together with the earlier OCM results \cite{Yasuro2}, the
agreement with experiment is quite satisfactory considering the
complexity of the task. With respect to the OCM result, from the
extended THSR calculation, one $0^+$ state is missing. This is due to
the fact that, so far negative parity states are not included in the
THSR scheme and then the state where an $\alpha$ is orbiting around
the $1^-$ state in $^{12}$C is missing. On the other hand, the $0^+_6$
state is well identified as the four $\alpha$ condensate state
analogous to the Hoyle state. This stems from the large radius of this
state which turns out to be rms = 4.9 fm, see also comment in
\cite{12C-dipole}. In Fig.\ref{Barbui} many more $\alpha$ states are
shown. But they are known states detected in earlier experiments, see,
e.g., \cite{Freer4}. We will not discuss those higher lying $\alpha$
cluster states here. So, this is about the status where we stand
concerning $\alpha$ gas spectral states in $^{16}$O.\\
One further recent development with respect to $^{16}$O relates to the
monopole spectral function. This has recently been developed by Yamada
{\it et al.} \cite{Yamada1}. In Fig.\ref{Yamada1,2} we show the
spectrum of the monopole transitions in $^{16}$O, once obtained from
the cluster states and once from the traditional RPA approach. We see
that both approaches are complementary. The cluster approach
reproduces the low lying part of the spectrum whereas the RPA rather
yields the higher lying part around the giant resonance region. The
sum-rule value of the cluster states amounts to $\sim$ 20\%.

\section{Cluster states in nuclei with neutron excess}

%%%%%%%%%%%%%%%%%%%%%%%%%%%%%%%%%%%%%%%%%%%%%%%%%%%%%%%%%%%%%

\begin{figure}[h]
%\begin{center}
\includegraphics[scale=0.46]{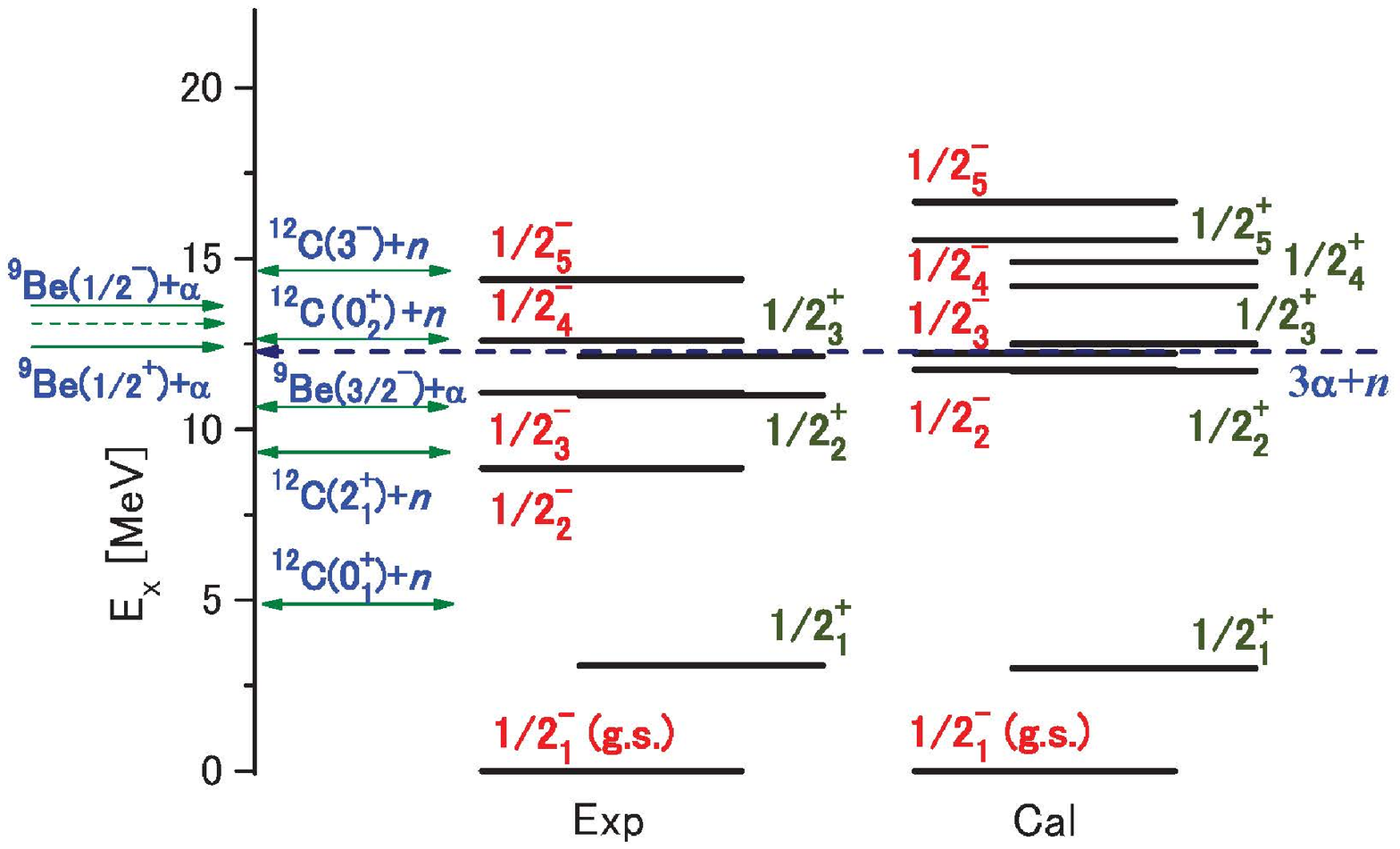}
\caption{Energy levels of the $1/2^-$ and $1/2^+$ states of $^{13}$C
  obtained from the 3$\alpha + n$ OCM calculation~\cite{Yamada2},
  compared with the experimental data. The threshold of the
  $^9$Be($5/2^-) + \alpha$ channel at $E_x = 13.1$ MeV, located
  between the $^9$Be($5/2^-) + \alpha$ and $^9$Be($1/2^-) + \alpha$
  channels, is presented by the dashed arrow on the left hand side of
  the panel. From \cite{Yamada2}.}
\label{fig:3a+n_levels}
%\end{center}
\end{figure}
%%%%%%%%%%%%%%%%%%%%%%%%%%%%%%%%%%%%%%%%%%%%%%%%%%%%%%%%%%%%%%

%%%%%%%%%%%%%%%%%%%%%%%%%%%%%%%%%%%%%%%%%%%%%%%%%%%%%%%%%%%%%
\begin{figure}
  \includegraphics[width=6cm]{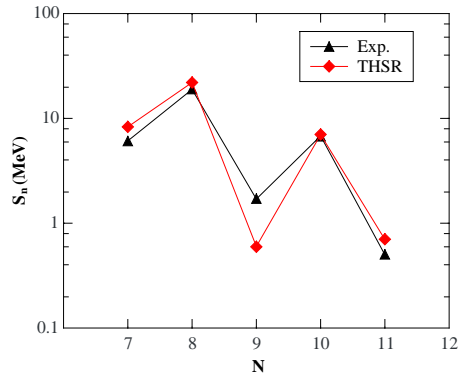}
  \caption{\label{valence-THSR}Result for the $^{8-11}$Be chain using
    a THSR wave function including valence neutrons. From \cite{Lyu}.}
  \end{figure}
%%%%%%%%%%%%%%%%%%%%%%%%%%%%%%%%%%%%%%%%%%%%%%%%%%%%%%%%%%%%%

So far, we only considered selfconjugate nuclei with N=Z. However,
$\alpha$ clustering should survive adding some extra neutrons. For
example is there some sign of a Hoyle like state in $^{13}$C or
$^{14}$C, etc. ? Yamada {\it et al.} \cite{Yamada2} have studied with
the OCM approach the spectrum of $^{13}$C, see
Fig. \ref{fig:3a+n_levels}. The state $(1/2^+)_5$ is clearly
identified as being composed by the Hoyle state plus an extra neutron
in an S-orbit \cite{Yamada2}. Also Lyu et al. \cite{Lyu} have recently
adapted the THSR approach to account for extra neutrons. In
Fig.\ref{valence-THSR} we show the neutron separation energies of the
chain $^{8-11}$Be as obtained by Lyu {\it et al.}. We also want to
mention an even more advanced work on $^9$B by Qing {\it et al.}
\cite{Qing} where in the THSR wave function additional proton-$\alpha$
correlations are incorporated.

\section{Alpha condensation seen as a Quantum Phase transition}

\begin{figure}
  \includegraphics[width=6cm]{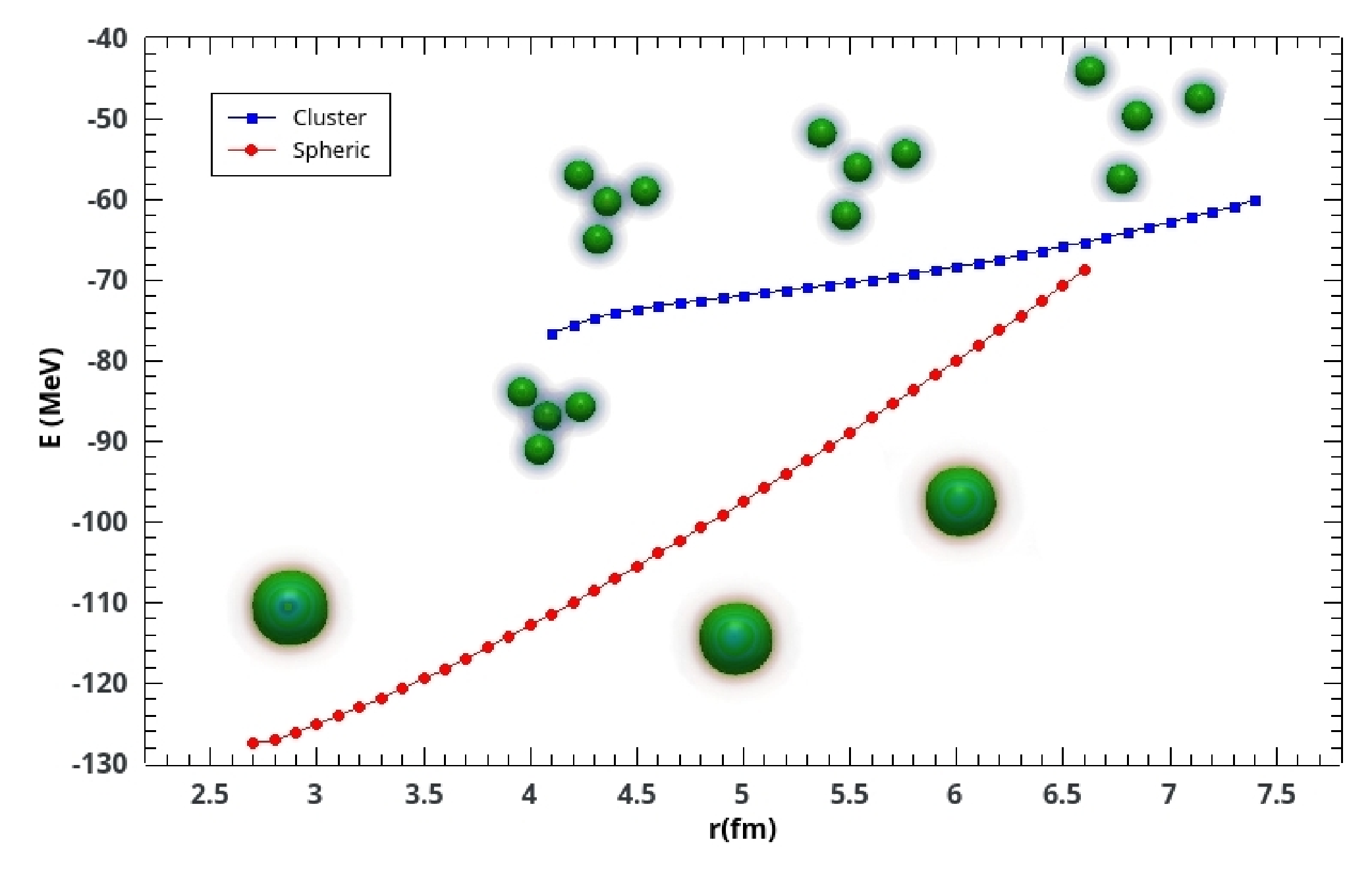}
\includegraphics[width=6cm]{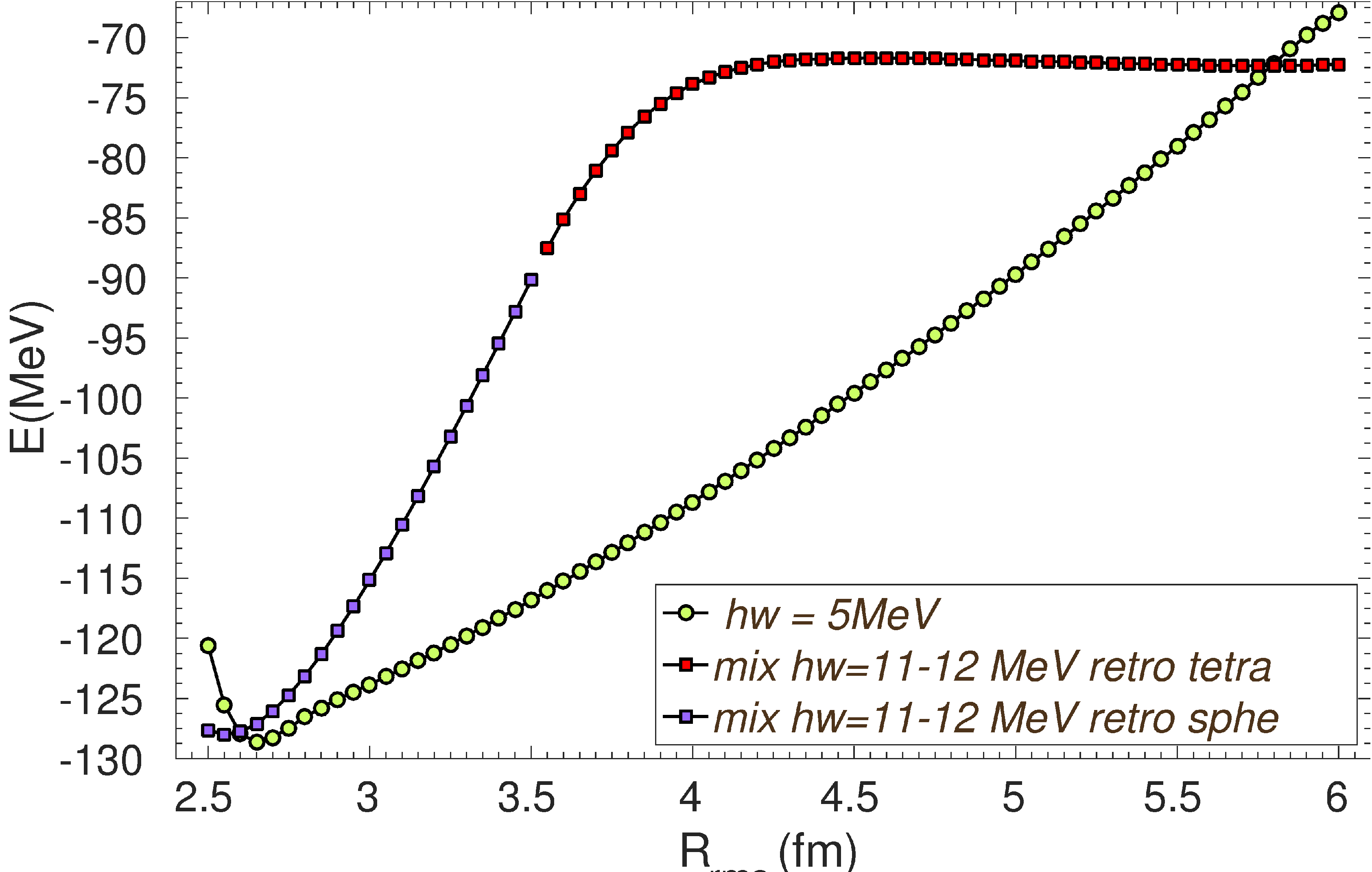}
\caption{\label{QPT} Physics of spherically extending $^{16}$O nucleus
  in HFB approximation. Left panel: with Gogny force (figure
  greatfully supplied by M. Girod); right panel: with RMF (figure
  greatfully supplied by J.P. Ebran).}
\end{figure}

\begin{figure}
\includegraphics[width=6cm]{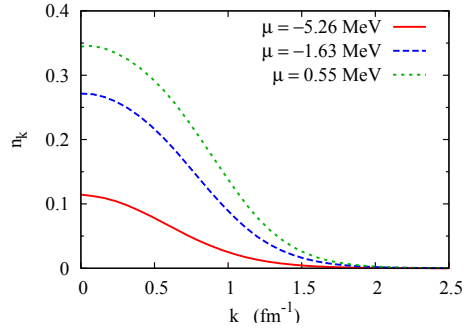}
\caption{\label{a-occs} Single particle momentum distribution in the $\alpha$ condensate for increasing values of chemical potential (density). At around zero chemical potential, $\alpha$ condensate disappears. From \cite{Sogo}.}
\end{figure}

Alpha condensation is a very nice example of a Quantum Phase
Transition (QPT) in nuclear physics \cite{Ebran1}. In Fig.\ref{QPT} we
show what happens to $^{16}$O in a HFB calculation under the
constraint that the radius is extended, that is density is lowered. In
the left panel the calculation has been performed with the Gogny force
by M. Girod and in the right panel the same is done employing RMF by
J.P. Ebran. The two figures are quite similar though there are
differences in detail. The lower branch represents a homogeneous
spherical expansion of the nucleus. On the upper branch the nucleus is
allowed to break up into a tetraeder of four $\alpha$ particles. In
both calculations there is a crossing of both branches. It means that
there is a transition from one phase to the other. This can be
qualified as a QPT where the density is the control parameter, the
system being at zero temperature. Not only there is a transition in
the spatial structure but also in the superfluid order parameter. As
long as the sphere is homogeneous pairing is non-zero. In the
tetraeder configuration pairing is strictly zero. It is interesting to
look into the details of the figure. One sees that the QPT occurs
quite a bit earlier with RMF than in the non-relativistic case. This
is in line with the observation made in earlier publication
\cite{Khan} that with RMF clustering is favored with respect to
non-relativistic functionals. This mean field calculations cannot be
quantitatively realistic, since the nucleus will not take on a
crystalline structure. However, the onset of $\alpha$ clustering is
reproduced correctly. In reality, the $\alpha$'s will be in a Bose
condensed state with a much lower energy. This has been studied in
infinite nuclear matter \cite{Sogo} at zero temperature. Also in this
study the condensate ends to exist beyond a quite small density, see
Fig.\ref{a-occs}, that is for a chemical potential around zero
equivalent to about a tenth of saturation density quite similar to the
transition density in $^{16}$O. In both studies the sudden
disappearance of $\alpha$ clustering is due to the Pauli
principle. Indeed, as density increases, antisymmetrisation
counteracts binding more and more rapidly and at the Mott-density
binding of the $\alpha$ particle is gone. This process has been
described at length in earlier publications \cite{Beyer}.

%%%%%%%%%%%%%%%%%%%%%%%%%%%%%%%%%%%%%%%%%%%%%%%%%%%%%%%%%%%%%%%%%%%%%%%%%%
%%%%%%%%%%%%%%%%%%%%%%%%%%%%%%%%%%%%%%%%%%%%%%%%%%%%%%%%%%%%%%%%%%%%%%%%

%%%%%%%%%%%%%%%%%%%%%%%%%%%%%%%%%%%%%%%%%%%%%%%%%%%%%%%%%%%%%%%%%%%%%%
%%%%%%%%%%%%%%%%%%%%%%%%%%%%%%%%%%%%%%%%%%%%%%%%%%%%%%%%%%%%%%%%%%%%%%%

\section{Physics of $\alpha$ preformation in $\alpha$ decay}

\begin{figure}
  \includegraphics[width=6cm]{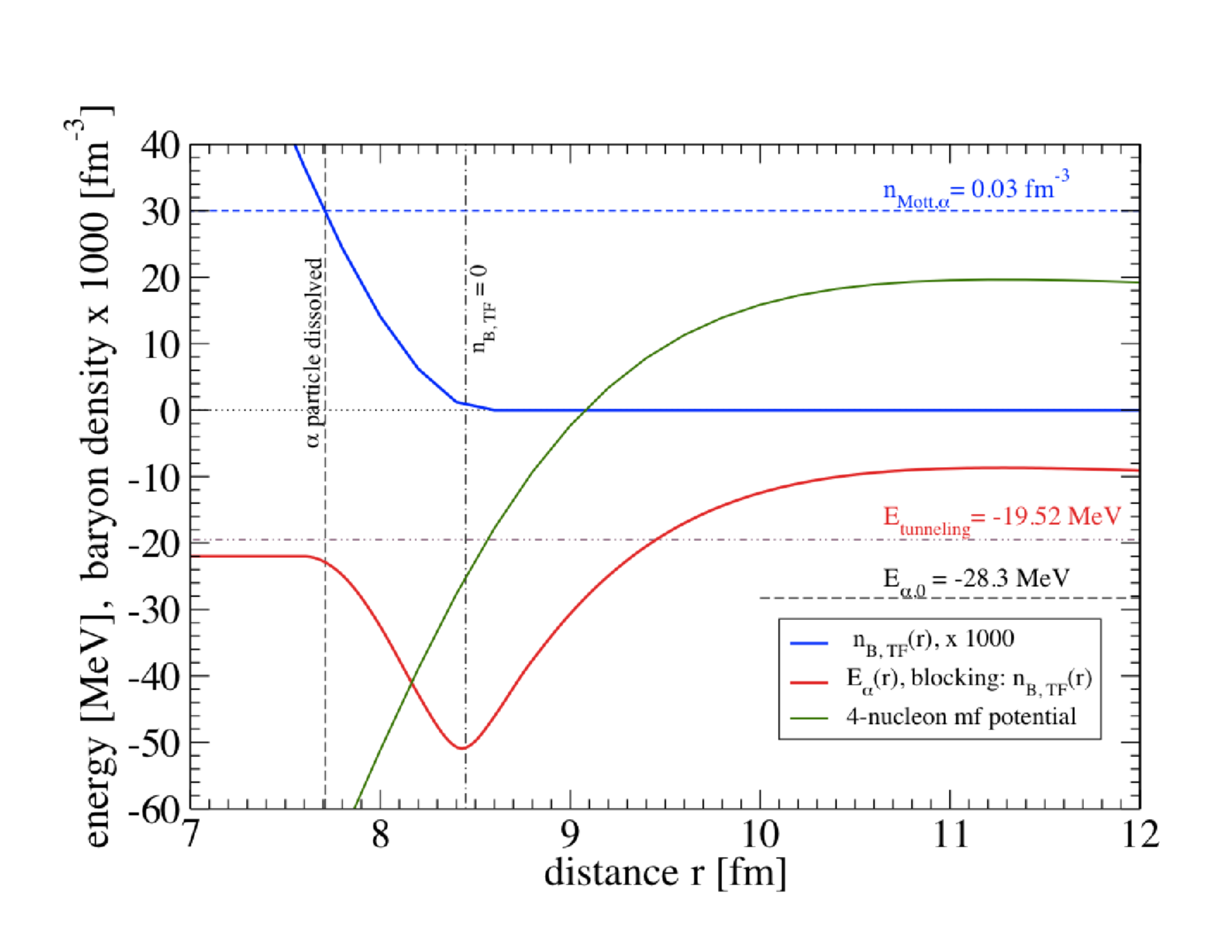}
  \includegraphics[width=6cm]{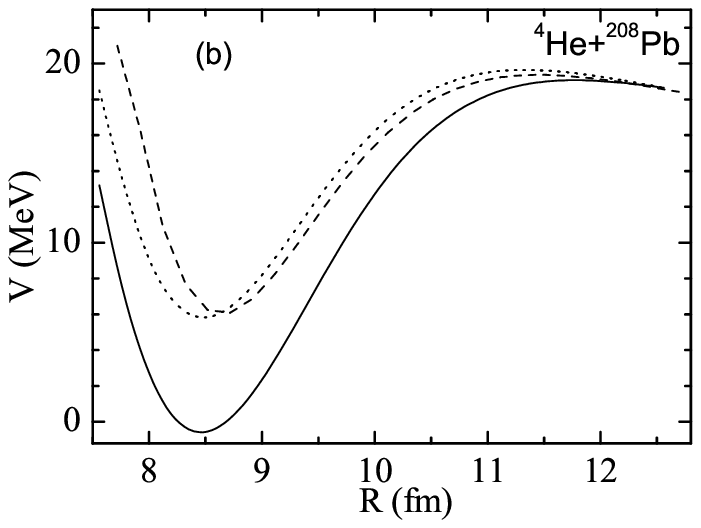}
  \caption{\label{Gerd} Formation of the $\alpha$ pocket on the
    surface of $^{208}$Pb in $^{212}$Po = $^{208}$Pb + $\alpha$ with
    the THSR approach (left, \cite{Gerd}) and the double folding
    approach (right, \cite{Adamian}).}
\end{figure}

The very strong sensitivity of $\alpha$ binding to the Pauli exclusion
principle with rising density is also at work in forming a potential
pocket for the $\alpha$ at the surface of, e.g., $^{208}$Pb
\cite{Gerd}. In a gedanken experiment, we imagine an $\alpha$ particle
approaching the $^{208}$Pb nucleus centrally. The four nucleons of the
$\alpha$ will first feel the attraction of the mean field of the lead
core. However, quickly the $\alpha$ particle will lose its binding in
coming closer to $^{208}$Pb. The potential energy of the $\alpha$
will, thus, go through a minimum before bending up again. Finally the
$\alpha$ will lose its identity fully, that is its intrinsic binding
is completely gone and the wave functions of the four nucleons turn
into shell model states. The complete microscopic description of this
process is very difficult. But at least the potential pocket of the
$\alpha$ can be made visible. In Fig.\ref{Gerd} we show such a pocket
behavior where the $\alpha$ is treated quantally whereas the
$^{208}$Pb core is considered in Local Density Approximation
(LDA). This result has again been produced with a THSR type of
approach. A quite different method which also produces a similar
pocket has been employed by Adamian et al. \cite{Adamian}. It consists
in the well known double folding procedure applied to the $^{212}$Po =
$^{208}$Pb + $\alpha$ system. The pocket is quite similar, see right
panel of Fig.\ref{Gerd}, to the THSR treatment. However, the
transition to the shell model regime is not well reproduced. It would
be interesting to study the relation between both approaches.

%%%%%%%%%%%%%%%%%%%%%%%%%%%%%%%%%%%%%%%%%%%%%%%%%%%%%%%%%%%%%%%%%%%
%%%%%%%%%%%%%%%%%%%%%%%%%%%%%%%%%%%%%%%%%%%%%%%%%%%%%%%%%%%%%%%%%%%%

\section{Conclusions}

In this contribution we have attempted to give an overview of recent
advances in the theoretical description of nuclear
clustering. Deliberately we restricted ourselves almost entirely to
$\alpha$ clusters. Naturally $\alpha$ gas states in $^{12}$C took up a
large part despite of the fact that great progress has been achieved
already since some while ago. However, we did not only point to
positive developments but also were critical with respect to some
recent results which we think are not on safe grounds or are at least
wrongly interpreted. We also pointed to missing investigations
concerning some approaches. This concerns for instance the
reproduction of the inelastic form factor of the Hoyle state what we
think is absolutely necessary to give a final credit to a correct
theory for the Hoyle state. We also emphasized the interpretation of
the Hoyle state as being to good accuracy an $\alpha$ condensate. This
stems from the fact that Ishikawa showed a link between the degree of
$\alpha$ particle condensation in $^{12}$C and relative S-wave motion
of $\alpha$'s. He found for both quantities a number of $\sim $ 80 \%
realisation. On the other hand there exist about half a dozen of more
works where a dominant relative S-wave motion of the $\alpha$'s had
been found underlining implicitly the condensate character. In this
respect it also would be desirable that more works attempt to
calculate the bosonic occupation factors of the $\alpha$'s in $\alpha$
gas states. So far only three results exist.\\
We went on to discuss the case of $^{16}$O. The situation there is
quite a bit more complicated than in $^{12}$C. In the latter nucleus,
it suffices to knock loose one $\alpha$ particle and automatically the
two others are also in a loosely bound configuration. In $^{16}$O on
the contrary, knocking one $\alpha$ loose means that the remaining
$^{12}$C nucleus has many possibilities to stay in a compact
configuration. According to our studies only the sixth $0^+$ state at
15.1 MeV can be identified as the first four $\alpha$ gas
state. Unfortunately the experimental situation in $^{16}$O is much
less advanced than for the $^{12}$C case. Most importantly the
measurement of the inelastic form factor, which gives so precious
information for the Hoyle state, is missing. But there is hope: I show
in this contribution in Fig.\ref{Barbui} an excitation spectrum
obtained from a coincident measurement of two $^8$Be (and $\alpha$ +
Hoyle state) obtained from the inverse kinematics reaction $^{20}$Ne
$+ \alpha$. A clear peak at 15.2 MeV can be seen. This might very well
be the Hoyle analogous state in $^{16}$O. It will be very exciting to
see more details on this in the near future.\\
Further important studies concern $\alpha$ clustering in nuclei with
valence neutrons. It is very likely that such states exist, e.g., in
$^{13}$C, $^{14}$C, etc. The works we cite show that the situation may
be more complex as assumed. Nevertheless, it seems that states of this
type can be identified in $^{13}$C.\\
A new development consists in the fact that $\alpha$ condensation can
be considered as one of the nicest examples in nuclear physics of a
Quantum Phase Transition (QPT). Indeed, the Pauli principle
counteracts condensation and as soon as the density of the system
reaches a point where the $\alpha$'s start to overlap substantially
with their density tails, very quickly the $\alpha$'s disappear at the
Mott density. So this QPT is triggered by the density as control
parameter.\\
The last issue is related to the former. If an $\alpha$
particle approaches a heavy nucleus as, e.g., $^{208}$Pb on a central
trajectory, the four nucleons will first feel the attraction of the
mean field of the lead nucleus. However, soon the Pauli principle will
come into action and the $\alpha$ will lose binding energy what
counteracts the gain of energy coming from the $^{208}$Pb mean
field. At the Mott density the $\alpha$ has disappeared as a cluster
and the four nucleons have gone over into standard shell model
states. In the end the whole process creates a potential energy pocket
at the surface, see left panel of Fig.\ref{Gerd}. The fully quantal
description of transition of the nucleon wave function in the $\alpha$
and to the shell model is very difficult. In Fig.\ref{Gerd} the lead
nucleus has been treated in Local Density Approximation (LDA) what
simplifies the calculation substantially. On the right panel of
Fig.\ref{Gerd}, we see a similar potential pocket obtained, however,
from a completely different approach. This is the well known double
folding procedure. The effective force of Fayans \cite{Fayans} was
employed which has sufficient repulsion at higher density to create
the pocket at the interior. It would be interesting to establish a
connection of the two approaches in the future.\\
I hope that with this short overview I was able to express my
excitement about $\alpha$ cluster physics and that I am spurring
further theoretical and experimental studies.

\section{Acknowledgements}

This short summary of recent advances in nuclear cluster physics is
based on intense work over the last two decades. This route was made
together in close collaboration with my colleagues, Y. Funaki,
H. Horiuchi, G. R\"opke, A. Tohsaki, T. Yamada. This contribution
could not have been written without their constant help and input over
the years. This is greatfully acknowledged. More recently, I had the
priviledge to also work with M. Lyu, Z. Ren, C. Xu, Bo Zhou, Qing
Zhao, all scientist from China. Their collaboration is very much
appreciated. Many thanks also to Takashi Abe for a very careful
reading of the manuscript and for providing Fig.7.


\begin{thebibliography}{1}
\bibitem{12C-dipole}
      If the $\alpha$ gas nature of the $1_1^-$ state in $^{12}$C is confirmed, see Fig. \ref{Kimura2}, then the state state where an $\alpha$ orbits around this state, i.e., the 5th $0^+$ state, could also be an $\alpha$ gas state.
\end{thebibliography}
\end{document}